# SANTA: Self-Aligned Nanotrench Ablation via Joule Heating for Probing Sub-20 nm Devices

Feng Xiong[1, *], Sanchit Deshmukh[1], Sungduk Hong[2], Yuan Dai[2], Ashkan Behnam[2], Feifei Lian[1], and Eric Pop[1, *]

[1]*Department of Electrical Engineering, Stanford University, Stanford, CA 94305, USA*
[2]*Department of Electrical and Computer Engineering, University of Illinois at Urbana-Champaign, Urbana, IL 61801, USA*

**ABSTRACT:** Manipulating materials at the nanometer scale is challenging, particularly if alignment with nanoscale electrodes is desired. Here we describe a lithography-free, self-aligned nanotrench ablation (SANTA) technique to create nanoscale "trenches" in a polymer like poly(methyl) methacrylate (PMMA). The nanotrenches are self-aligned with carbon nanotube (CNT) and graphene ribbon electrodes through a simple Joule heating process. Using simulations and experiments we investigate how the Joule power, ambient temperature, PMMA thickness, and substrate properties can improve the spatial resolution of this technique. We achieve sub-20 nm nanotrenches for the first time, by lowering the ambient temperature and reducing the PMMA thickness. We also demonstrate a functioning nanoscale resistive memory (RRAM) bit self-aligned with a CNT control device, achieved through the SANTA approach. This technique provides an elegant and inexpensive method to probe nanoscale devices using self-aligned electrodes, without the use of conventional alignment or lithography steps.

[*]Contact: epop@stanford.edu

# 1    Introduction

Much interest and active research exists in one-dimensional (1D) materials such as nanowires [1, 2] (NWs), carbon nanotubes (CNTs) [3, 4] and graphene nanoribbons (GNRs) [5-7] due to their interesting mechanical, electrical and thermal properties compared to those of bulk materials. CNTs and GNRs have specifically attracted attention for future electronic applications due to their nanoscale dimensions and high current carrying capabilities [8]. CNTs and GNRs could also be used as extremely sharp nanoscale electrodes for other nanomaterials such as molecules [9], DNA [10] or memory bits [11]. In order to probe such nanomaterials, it is crucial to align the CNT or GNR with the device being probed, and to achieve this with nanoscale resolution. Although electron-beam (e-beam) lithography could, in principle, be used to create patterns with nanometer scale spatial resolution, it is time-consuming, expensive, and alignment to individual CNTs (1-2 nm diameter) would be extremely challenging.

In previous work, we proposed a simple mechanism for utilizing Joule heating from CNTs in order to pattern nanoscale "trenches" in a polymer [12]. Similar lithography-free techniques have also been proposed for sensor applications [13], study of biological and chemical phenomena [14], catalytic NW growth [15], and selective removal of metallic CNTs [16]. Localized heating has also been used as a simple patterning approach for nanoscale positioning [17-19]. Microheaters have been studied for selective functionalization of sensors [20] and catalytic synthesis of nanomaterials [18]. The advantage of using CNTs as Joule patterning electrodes is that given their small diameter (1-2 nm), nanoscale patterns of this size ought to be achievable. However, only patterns of ~50 nm widths were achieved in previous work, and an understanding for the creation and optimization of such nanoscale features by Joule heating has been lacking until now.

In this work, we push the Joule patterning method towards ~10 nm scale features, achieved through a deeper understanding and optimization of nanoscale heating. We first present a three-dimensional (3D) finite element method (FEM) model of Joule heating around a CNT covered by a polymer film. This provides a predictive platform revealing all variables controlling the formation of nanoscale features during heating. Guided by these simulations, we fabricate sub-20 nm nanotrenches in polymethyl methacrylate (PMMA) by tuning the input power, substrate temperature and PMMA thickness. We also report successful nanopatterning with narrow two-dimensional (2D) graphene heaters, down to ~30 nm width for the first time. Unlike conventional lithography techniques, the self-aligned nanotrench ablation (SANTA) technique automatically aligns the devices to be tested with the 1D or 2D probing electrodes. As a novel test case, we report a working nanoscale resistive random access memory (RRAM) bit self-aligned with a CNT control device, entirely fabricated through the SANTA approach.

## 2 Results and discussion

### 2.1 Process of forming nanotrenches

Figures 1 demonstrates our SANTA nanotrench formation process and the atomic force microscopy (AFM) images of the CNT and the nanotrench. We start with a CNT device with Pd electrodes on a $SiO_2/Si$ substrate. CNT growth and device fabrication steps were reported in detail elsewhere [21]. CNTs with different chiralities (metallic or semiconducting) and different diameters (single-walled or small diameter multi-walled) could all be used in the SANTA technique, as long as they pass a sufficiently high current to heat and pattern the PMMA (e.g. semiconducting CNTs need be gated in the "on" state, while metallic CNTs can be used without gating). We spin coat a thin layer of PMMA on top of the CNT, with thickness varying from 20 to 60 nm, controlled by the spin rate, see electronic supplementary material (ESM). We then apply a constant voltage across the electrodes to induce current flow in the CNT. Joule heating from the CNT heats up the PMMA film (Fig. 1a) causing it to ablate, leaving behind a nanotrench self-aligned with the CNT (Fig. 1b, d). Insets in Figs. 1a, b show the cross-sectional view perpendicular to the CNT. The Fig. 1d inset shows the depth profile of the resulting nanotrench, with the CNT at its bottom.

We note that if the input power is insufficiently high, the nanotrench may not form along the CNT all the way to the two electrodes, due to heat sinking from the metal contacts [22, 23]. We perform the entire process in an environmentally controlled probe station (typically under ~$10^{-5}$ Torr vacuum) to prevent CNT breakdown and control the substrate temperature. The resulting nanotrench provides a convenient platform to position active materials such as molecules for sensor and biological studies [9, 10, 14, 24], dielectric nanowires or memory bits [11, 12], synthesis of nanowires via catalytic growth or direct deposition [15, 18, 19], and removal of unwanted CNTs for sorting after plasma etching [16].

### 2.2 Simulation platform

To better understand the heating and temperature distribution in our devices, we developed a 3D COMSOL FEM model that is consistent with our experimental setup, as shown in Fig. 2a. To improve the efficiency, only a quarter of the actual device is simulated, taking advantage of the two symmetry planes. Figure 2a shows the first symmetry plane bisects the CNT into two half-cylinders along its axis; while the second symmetry plane bisects the CNT in the middle, perpendicular to its axis. The second assumption is valid for metallic CNTs (which have uniform heating) and is a reasonable approximation for semiconducting CNTs under high bias [21]. The size of the simulated Si substrate is $20\times20\times20$ $\mu m^3$, sufficiently large to capture all heat spreading away from the CNT, but small enough to facilitate meshing and computation [25]. The bottom surface and the two side surfaces (non-symmetry planes) are held at the ambient temperature (isothermal boundary conditions), while all other outer surfaces are treated as thermally insulating (adiabatic boundary conditions).

Figure 2b shows a zoomed-in image of the CNT heater, and confirming that the greatest temperature gradients are very close to the CNT itself. The 3D model includes thermal boundary resistances (TBR) at all interfaces, matched against data from the literature. The lumped TBR at the CNT-Pd interface is $R_{th,c} \approx 1.2 \times 10^7$ K/W and a thermal boundary conductance $g = 0.17$ W/K/m (per CNT length) is applied at the CNT-SiO$_2$ boundary [26]. All other interior interfaces (Si-SiO$_2$, SiO$_2$-PMMA, and CNT-PMMA) have $R_{th} = 2.5 \times 10^{-8}$ m$^2$K/W per unit area [7, 25, 27]. The thermal conductivities of the CNT, SiO$_2$, Si, PMMA and Pd are taken as 2200, 1.4, 150, 0.1 and 70 W/K/m, respectively [7, 22]. The thermal model solves the steady-state 3D heat diffusion equation to obtain the temperature profile in the device. (The heat diffusion equation is appropriate without any ballistic corrections because the CNT is long and the materials immediately surrounding it are amorphous, with low thermal conductivity.) Figure 2c shows a typical cross-sectional temperature profile in the middle of the CNT (same orientation as the inset of Fig. 1b). The nanotrench is formed where the PMMA temperature exceeds its boiling point (~523 K), as shown by the void (white) region in Fig. 2c. The inset shows the sharp temperature profile along the lateral direction with a peak temperature gradient of ~2.7 K/nm. This confirms that the CNT heater delivers highly localized temperature profile and thus form nanoscale trench.

## 2.3 CNT heaters

As our next step, in Figs. 3 and 4 we systematically study the effect of input power per CNT length ($P_L$), ambient temperature ($T_0$), PMMA thickness ($t_{PMMA}$) and substrate type on the width of the nanotrench ($W$), defined as the narrowest region where all PMMA is heated above its boiling temperature. Figure 3a shows how the nanotrench width $W$ varies as a function of the input power per length $P_L$, at different ambient temperatures. Here, the PMMA thickness $t_{PMMA} = 40$ nm and the CNT is on 90 nm thick SiO$_2$, on a 500 $\mu m$ thick Si wafer, consistent with our initial experiments.

Dashed lines are simulation results from FEM and symbols are the measured experimental results. The nanotrench widths were measured by AFM using sharp single crystalline diamond (SCD) probes with a tip radius ~5 nm, after accounting for the tip convolution effect [28]. We varied the temperature in steps of 50 K from 150 K to 350 K for various CNTs otherwise treated under similar conditions. Each temperature is color coded as shown in the diagram.

Our simulations generally agree with the experimental data, and most importantly the *trends* are confirmed: the nanotrench $W$ increases sub-linearly (almost logarithmically) with $P_L = I(V-IR_C)/L$, which excludes the voltage drops at the contacts. Here we estimated the electrical contact resistance $R_C$ from the CNT resistance at low voltage, when $R_C$ is expected to dominate for our process conditions [21]. The normalized Joule heating power per length $P_L$ is a better parameter to quantify the ablation process than either applied voltage $V$ or current $I$ alone, since each CNT device may have different resistance, chirality and channel length. While we could use either constant voltage or constant current approach to achieve the desired input power density, we chose the constant voltage method in this work to limit the voltage across the CNT and avoid device breakdown. The sublinear dependence of $W$ on $P_L$ is not unexpected, because as the nanotrench grows in width, more heat (from the CNT) will be lost to the underlying substrate rather than heating up the PMMA. We also see that the power required to form a nanotrench with a given width decreases with increasing ambient temperature, as it is easier to heat up and evaporate PMMA at elevated temperatures. Thus, heating the substrate may be desired in applications when semiconducting CNTs cannot generate enough power to "burn" through a relatively thick layer of PMMA. At lower temperature, nanotrench widths are naturally smaller for a given $P_L$, allowing us to achieve sub-20 nm wide nanotrenches at 150 K ambient. Qualitatively, at lower temperature, only PMMA that is in close proximity to the CNT will be heated up to sufficient temperature, thus producing narrower nanotrenches. At the same time, the viscosity of a liquid is generally lower at lower temperature, limiting the reflow of the film and assisting to keep the width small.

We next examine the effect of PMMA thickness on CNT nanotrench width, as shown in Fig. 3b. In this case, the ambient is 300 K and the substrate is 90 nm $SiO_2$. We control the PMMA thickness by varying the spin rate and/or adding A-thinner into PMMA. We use the ellipsometer to measure the PMMA thickness, and we include tabulated results of PMMA thickness as a function of spin rate and A-thinner concentration in the ESM. With thinner PMMA, the minimal achievable nanotrench width is smaller, as the minimal power to evaporate a thinner layer of PMMA is lower. From the simulation (Fig. 3b), we also observed that the rate of increase of $W$ with respect to the input power $P_L$ is smaller for thinner films. This is beneficial, as it allows us to have better control over the nanotrench width and better tolerance of variation in input power. However, for applications where a lift-off process is necessary (depositing dielectrics, memory bits, or metal nanowires [12]), we cannot reduce the PMMA thickness below a certain value [29], depending on the deposited material type and thickness.

We also look at how different substrates may affect the temperature distribution and thus the nanotrench width around CNT heaters. In Fig. 3c, we vary the $SiO_2$ thickness $t_{ox}$ = 30, 90 and 300 nm, and also look at the temperature profile if the CNT sits on quartz substrate. The ambient temperature is 300 K and the PMMA thickness is 40 nm. In the typical $SiO_2$/Si configuration, we see that with thinner $SiO_2$ the nanotrench width is significantly lower at the same input power. With thinner oxide, "vertical" heat dissipation from the CNT through the oxide to the thermally conductive Si substrate is easier. As a result, there is less lateral heat spreading from the CNT heater to the PMMA, resulting in sub 10-nm nanotrench with 30 nm of $SiO_2$ (note the logarithmic vertical scale in Fig. 3c). However, if the CNT is placed on a quartz substrate (thermal conductivity ~10 W/m/K), lateral heat dissipation is significantly larger. The trench width thus increases almost exponentially with $P_L$, because the entire quartz substrate is heated up and therefore heats up the PMMA from beneath (see ESM Fig. S3). Conversely, we note that a more unusual, thermally anisotropic substrate such as $La_5Ca_9Cu_{24}O_{41}$ (LCCO), which is thermally conductive in the vertical direction (~100 W/m/K) but insulating in the lateral direction (~2 W/m/K) [30], could be an ideal choice for creating ultra-narrow trenches.

It is interesting to note that the nanotrench may take up to a few seconds to reach its steady-state width, as shown in Fig. S4 of the ESM. These times are much longer than the thermal time constants of such a system (hundreds of nanoseconds) [27, 31], suggesting PMMA may reflow during this process. While our simulations cannot account for the complicated evaporation, melting and reflow processes, they nevertheless provide good agreement with the experimental data (Fig. 3). More importantly, these steady-state simulation results illustrate clearly how different parameters (input power density, ambient temperature, PMMA thickness and substrate thermal resistance) affect the SAN-TA process and allow us to fine-tune these parameters to achieve the desired nanotrench width for different applications.

As stated earlier, these PMMA nanotrenches can be filled with other materials, such as evaporated metals. After PMMA lift-off, this will then lead to the formation of a metal nanowire that is self-aligned with the CNT underneath. In Fig. 4a, we show such a typical metal nanowire, generated from the nanotrench after metallization (Cr, 5 nm) and lift-off. Here the trench was formed by applying an input power $P_L \approx 0.47$ mW/μm along the CNT at 300 K. The CNT heater was on a 30-nm SiO$_2$ (on Si) substrate and was covered with 40-nm of PMMA. Scanning electron microscope (SEM) image confirmed uniform coverage of the metal nanowire along the CNT. The inset shows a zoomed-in high-resolution SEM image revealing that the width of this nanowire to be ~20 nm, consistent with our simulation predictions.

## 2.4  CNT-Based Memory Devices

Because of the contact cooling effect, the temperature of the CNT which forms the nanotrench is lower near the metal contacts, over a distance comparable to the so-called thermal healing length along the CNT, here $L_H \approx 200$ nm [22, 23, 32]. Thus, depending on input power, a nanotrench may not always form all the way to the metal contact, enabling the existence of a short CNT portion in series with the nanowire formed in the nanotrench.

Here we exploited this capability to demonstrate a novel device: a CNT built-in series resistor and selection device for nanoscale resistive random access memory (RRAM). By controlling the SANTA input power, we limited the nanotrench in the PMMA from reaching the contact. We then evaporated 5 nm of Cr followed *in situ* by 5 nm of AlO$_x$, without breaking vacuum. Performing PMMA lift-off, this yields a CNT-Cr-AlO$_x$ coaxial nanowire in series with a short CNT segment (protected by the intact PMMA near the contact), as shown in Fig. 4b. The short CNT segment could serve the role of current compliance resistor (if metallic) or selection device (if semiconducting) in RRAM applications, without additional fabrication steps. To enable electrical measurements, we finally patterned a top metal electrode (Ti/Pd, 2/30 nm) to form a crossbar RRAM device, as illustrated in the SEM image (Fig. 4c) and the schematics (Fig. 4d inset). As the applied electric field increases across the oxide, a filamental conduction path forms in the otherwise insulating AlO$_x$, resulting in a significant drop of the device resistance. In typical RRAM devices, this sudden drop in resistance may lead to large increase in current density and possible thermal runaway. In this context, the built-in CNT segment acts as a current compliance resistor, as shown in the measured *I-V* characteristics (Fig. 4d).

## 2.5  GNR heaters

In addition to the 1D CNTs, we also used graphene nanoribbons (GNRs) with widths varying from 30 nm to 400 nm, as heaters to demonstrate the broad usefulness of the SANTA technique. Detailed fabrications of GNRs are reported elsewhere [33]. In short, we grow single-layer graphene on Cu foils by chemical vapor deposition (CVD), transfer them onto 90 nm $SiO_2$ substrates (on Si), and pattern them into ribbons by e-beam lithography. The electrode pads are Ti/Au (0.5/30 nm). A schematic of a typical GNR device is shown in the Fig. 5b inset. Figure 5 shows experimental (dots) and simulation (dashed lines) results of GNR nanotrench formation, with different GNR widths (30, 60, 200 and 400 nm). The PMMA thickness and $SiO_2$ thickness are 40 and 90 nm, respectively.

As the heater dimension goes from 1D (CNTs) to quasi-2D (GNRs), the general trends still hold, i.e. the nanotrench width grows sub-linearly with the input power and the nanotrench width is smaller at lower temperatures. As the GNR heater width increases, the minimal achievable nanotrench width increases accordingly, being comparable to the GNR width. This suggests that our SANTA technique could be extended to 2D materials such as graphene and transition metal dichalcogenides (TMDs) for patterning, functionalization and sensor applications. (Recent studies have shown up to 300 µA/µm current density in $MoS_2$ [34], yielding $P_A \sim 20$ mW/µm² which is sufficient to remove the PMMA as seen in Fig. 5.) We plot all our experimental results (CNTs and GNRs) at $T_0 = 300$ K, $t_{PMMA} = 40$ nm, $t_{ox} = 90$ nm in Fig. 6. The horizontal axis is the power density per unit contact area with the substrate ($P_A$) and the vertical axis is the nanotrench width $W$ normalized by the width of the device under test ($W_{DUT}$), i.e. the CNT diameter or GNR width. The dashed line is a power law fitting with exponent ~0.943 (see inset), consistent with the sub-linear trend we described earlier.

## 3    Conclusions

In summary, we described a self-aligned nanotrench ablation (SANTA) technique, using Joule heating of a nanoscale heater (e.g. CNT or GNR) to create nanotrenches in PMMA along the direction of the underlying CNT or GNR. These nanotrenches can be used to self-align materials or devices to be probed at ~10 nm scales with the CNT or GNR electrode. We developed a 3D simulation model to understand and optimize the trench formation process, showing that with lower ambient temperatures, thinner PMMA and substrate oxide, we gain better control over the nanotrench width. Guided by these simulations, we also experimentally demonstrated sub-20 nm nanotrenches at 150 K, with 30 nm PMMA film and using CNT heaters. Sub-10 nm nanotrenches could be achieved by using thinner substrate oxide thicknesses for better "vertical" heat sinking. This technique could be extended to other nanoscale electrodes (e.g. nanowires or 2D materials), providing a convenient way to position active materials with nanoscale precision. We demonstrate one potential application, building a nanoscale resistive random access memory (RRAM) bit self-aligned with a built-in CNT series resistor and selection device. Other applications include dielectric or metallic nanowires, selec-tively functionalizing sensors, facilitating catalytic nanowire growth, and removing undesired CNT connections.

## Acknowledgements

We thank Dr. Eilam Yalon and Dr. Ilya Karpov for technical support and helpful discussions. We acknowledge partial support from the National Science Foundation (NSF) CAREER grant 1430530, SRC / Intel grant 2014-IN-2532, the Stanford SystemX Alliance, and the Stanford Nano- and Quantum Science and Engineering (NQSE) Postdoctoral Fellowship (F.X.).

**Figures**

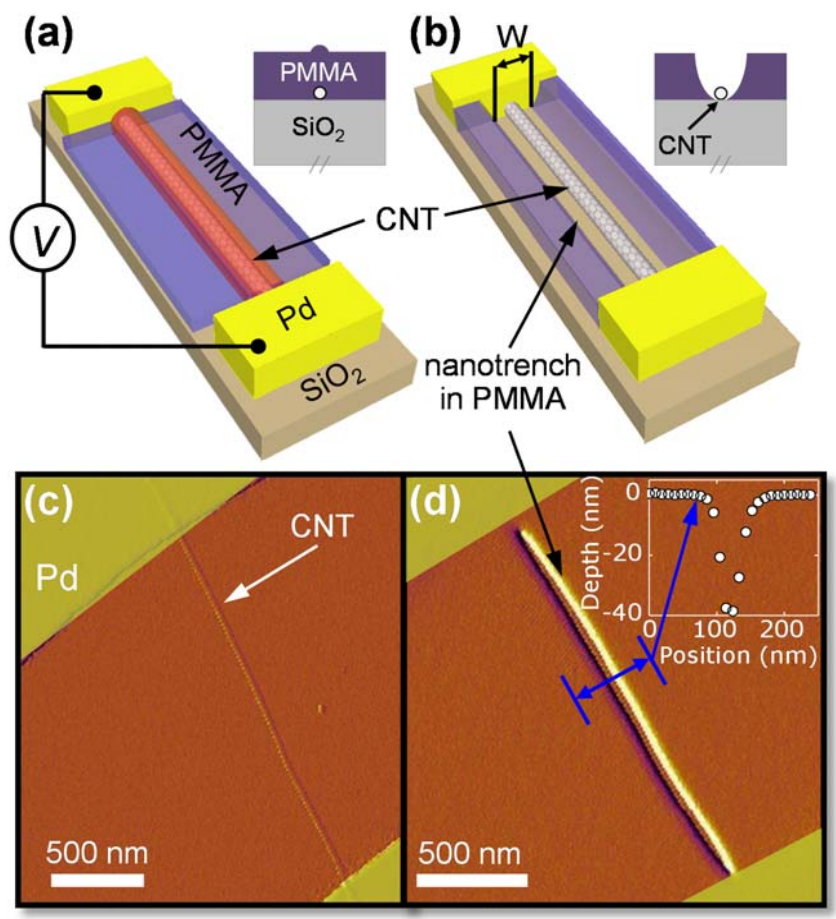

**Figure 1** The self-aligned nanotrench ablation (SANTA) technique: (a) Joule heating in a carbon nanotube (CNT) leads to nanotrench formation in a PMMA covering the CNT, as the polymer evaporates. (b) The nanotrench is self-aligned with the CNT heater. Insets show cross-sectional views. (c) False-color atomic force microscope (AFM) image of a CNT before PMMA deposition. (d) AFM image of the nanotrench formed in PMMA. Inset shows depth profile across the nanotrench measured by AFM.

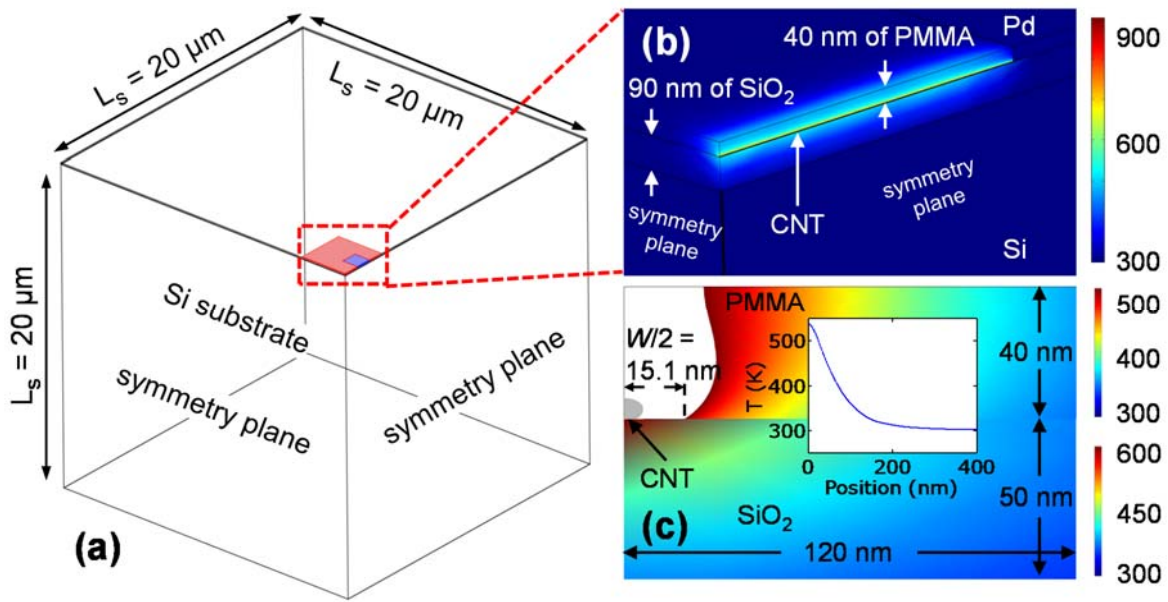

**Figure 2** (a) Schematics of 3D heating model of the test structure. Only a quarter of the actual device is simulated, taking advantage of symmetry conditions. The highlighted region is where the CNT heater is located. (b) Temperature profile in the device due to Joule heating from the CNT heater when $T_0 = 300$ K. (c) Cross-sectional temperature profile at the middle of the CNT. The white region represents the PMMA volume whose temperature exceeds the evaporation of PMMA (~573 K). The nanotrench width $W$ is defined as the narrowest region where all PMMA is heated above its boiling temperature. The CNT is drawn disproportionally large for clarity and two different temperature scales are used to better illustrate the temperature distribution in upper (PMMA) and lower (SiO$_2$) regions, respectively. The inset shows the temperature profile away from the CNT heater, with a maximum temperature gradient of ~2.7 K/nm, similar to the temperature gradient of the nanowire heater reported by Jin *et al.* [18].

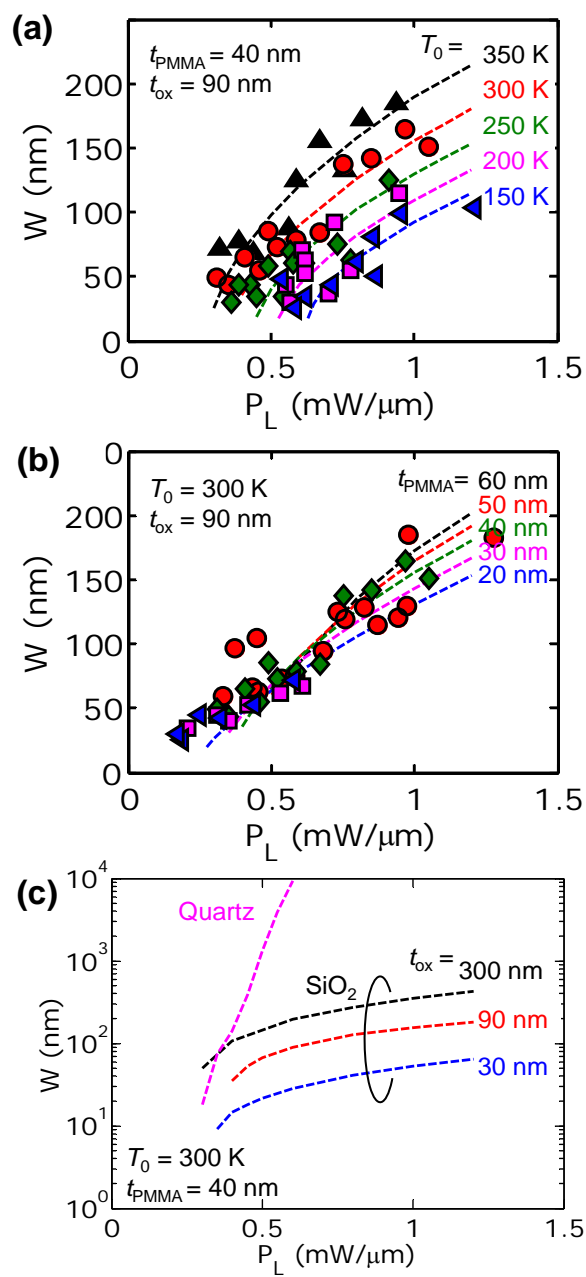

**Figure 3** Nanotrench width in PMMA as a function of input power of the CNT heater. Dashed lines are simulation results from the COMSOL model. Symbols are experimental data. (a) Effect of ambient temperature on nanotrench width. PMMA thickness is 40 nm and substrate is 90 nm SiO₂. (b) Effect of PMMA thickness on nanotrench width. Ambient temperature is 300 K and the substrate is 90 nm SiO₂. (c) Effect of substrate on nanotrench width. Ambient temperature is 300 K and PMMA thickness is 40 nm.

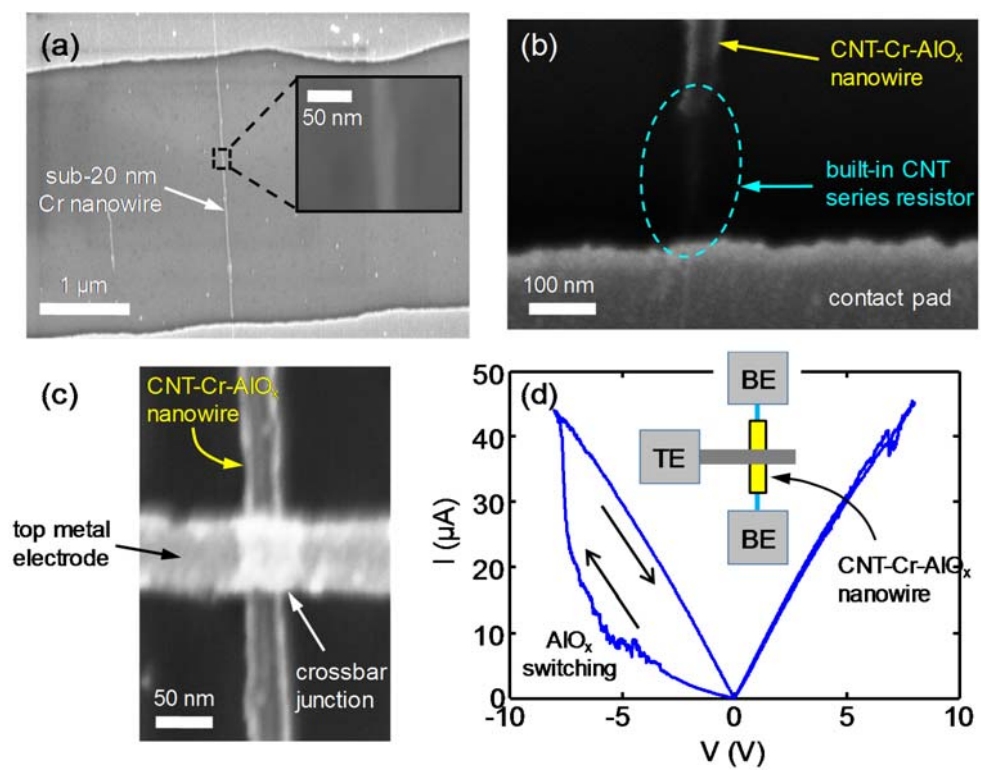

**Figure 4** Building co-axial nanowires and RRAM devices with the SANTA technique. (a) SEM image of a Cr nanowire formed along the CNT after metal evaporation and PMMA lift-off, showing uniform metal coverage of the CNT. Inset is a zoomed-in high resolution SEM revealing ~20 nm nanowire width. (b) SEM image of CNT-Cr-$AlO_x$ nanowire spanning only the middle region of the CNT. (c) SEM image of a crossbar memory (RRAM) device with the $AlO_x$ as the resistive switching material. The short uncovered CNT in Fig. 4b serves as a built-in series resistor and a selection device. (d) Measured current-voltage of the crossbar device. The inset is the schematic of the crossbar RRAM, where the CNT serves as the bottom electrode (BE). The top electrode (TE) is separately patterned.

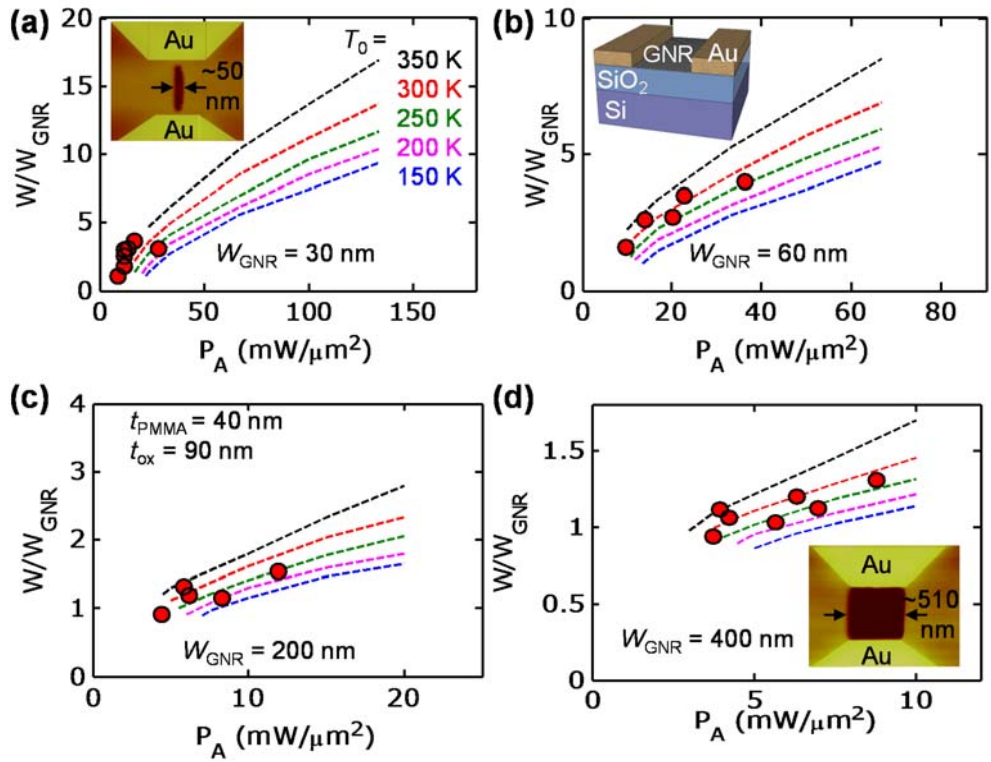

**Figure 5** Patterning of nanotrenches into PMMA using graphene nanoribbon (GNR) heaters. Plots show normalized nanotrench width by the GNR width as a function of input power per area ($P_A$) in GNRs at different ambient temperatures (150 K to 350 K from bottom to top). PMMA thickness is 40 nm and substrate is 90 nm SiO$_2$. Dashed lines are COMSOL simulation results and symbols are experimental data at 300 K. (a-d) GNRs with width of 30, 60, 200 and 400 nm, respectively. Insets in (a, d) are false-color AFM images of nanotrenches in PMMA after Joule heating in GNRs. (b) inset shows the schematic of a GNR device.

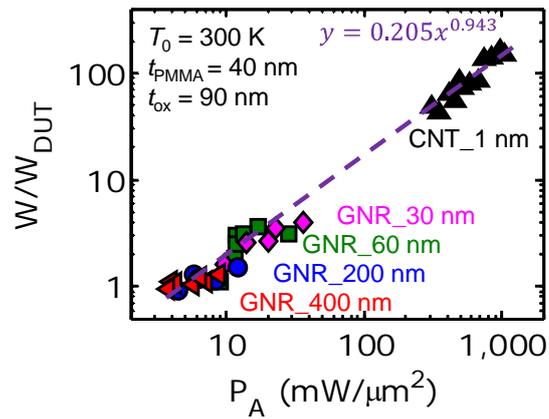

**Figure 6** Normalized nanotrench width by the width of the device under test (i.e. GNR width or CNT diameter) as a function of input power density (per unit area). Temperature is 300 K, PMMA thickness is 40 nm and substrate is 90 nm SiO₂. Symbols are experimental results, the dashed line shows a power law fitting with exponent of 0.943, consistent with the sub-linear trend we observed between nanotrench width and input power.

**Graphical table of content**

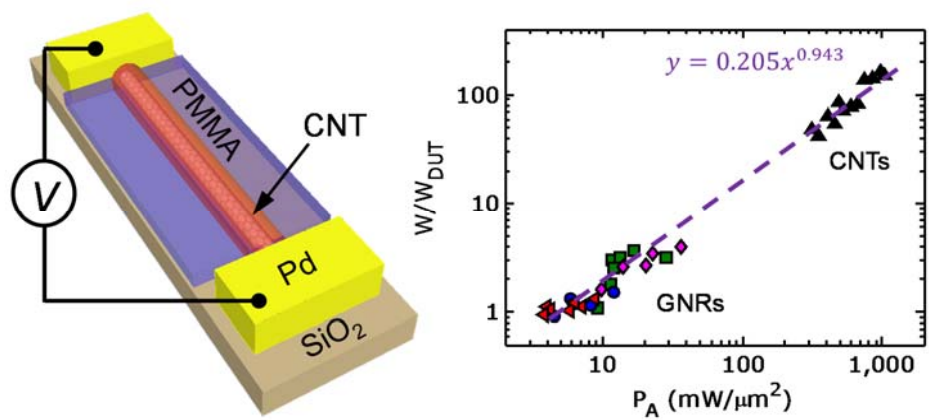

Nanotrenches are formed in PMMA by carbon nanotubes (CNT) and graphene nanoribbons (GNRs) via Joule heating. The nanotrench width can be sub-20 nm under optimized conditions, enabling probing of nanomaterials with self-aligned CNT or GNR electrodes.

## Electronic Supplementary Material

## SANTA: Self-Aligned Nanotrench Ablation via Joule Heating for Probing Sub-20 nm Devices

Feng Xiong[1], Sungduk Hong[2], Yuan Dai[2], Ashkan Behnam,[2] Sanchit Deshmukh[1], Feifei Lian[1] and Eric Pop[1]

[1]Department of Electrical Engineering, Stanford University, Stanford, CA 94305, USA
[2]Department of Electrical & Computer Engineering, University of Illinois at Urbana-Champaign, Urbana, IL 61801, USA

_**PMMA Spin Coating.**_ We vary the PMMA thickness by controlling the spin rate (RPM) and the film composition – adding A-thinner to dilute the film.

| Solution Composition | RPM | Thickness (nm) |
|---|---|---|
| 30% PMMA, 70% A-thinner | 4000 | 18 |
| 50% PMMA, 50% A-thinner | 4000 | 27 |
| 75% PMMA, 25% A-thinner | 4000 | 40 |
| 100% PMMA | 6000 | 45 |
| 100% PMMA | 4500 | 52 |
| 100% PMMA | 3500 | 57 |
| 100% PMMA | 2500 | 86 |
| 100% PMMA | 1500 | 113 |

**Table S1** PMMA thickness. PMMA thickness under different spin rates and compositions.

_**Nanotrench at Low Temperature.**_ The nanotrench width ($W$) is smaller at lower ambient temperature if all other conditions are the same. A comparison of two simulated nanotrench profiles at different ambient temperature (150 and 300 K, respectively) is shown in Fig. S2 below.

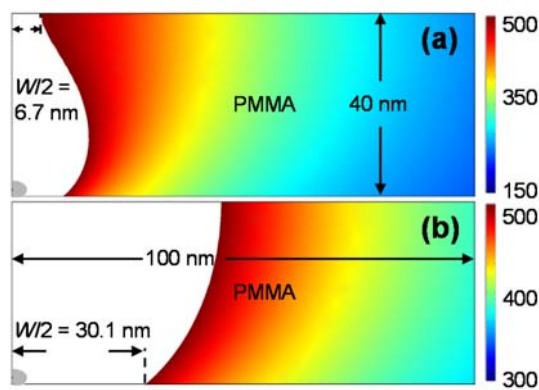

**Figure S2** Nanotrench formed by CNT at different ambient temperatures. (a) $T_0 = 150$ K. (b) $T_0 = 300$ K. The PMMA thickness is 40 nm and the substrate is 90 nm $SiO_2$ on a Si wafer (not shown).

***Quartz Substrate.*** We notice that if we perform trench formation simulations on quartz substrates, the trench width increases strongly (super-linearly) with the input power. We investigate this further with finite element model. We build a model with the full thickness of the device (500 μm) instead of the reduced size (20 μm). We compare the substrate temperature profiles for devices on quartz and on 90 nm $SiO_2$ on Si in Fig. S3. We notice that due to the presence of the thermally resistive $SiO_2$ ($k_{SiO2}$ ~ 1.4 W/m/K) layer, the heating in the substrate is very localized (Fig. S3a). With the more thermally conductive quartz ($k_{quartz}$ ~ 10 W/m/K) substrate, the entire substrate is being heated up (Fig. 3b). This heats up the PMMA film and creates much wider trenches.

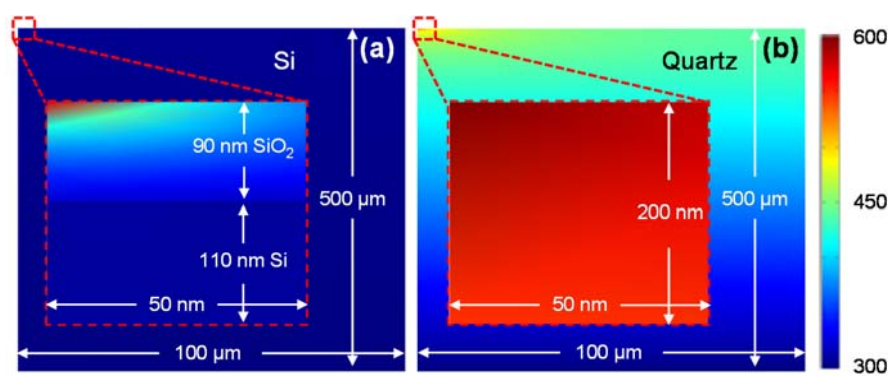

**Figure S3** Substrate temperature profile when the CNT power is 0.5 mW/μm. PMMA thickness is 40 nm and ambient temperature is 300 K. The CNT is at the top left corner. (a) Substrate is 90 nm $SiO_2$ on 500 μm Si. (b) Substrate is 500 μm quartz. The higher thermal conductivity of quartz broadens the lateral temperature profile and significantly widens the PMMA trenches near the CNT.

***Time Dependence.*** In Fig. S4, we measure the nanotrench width formation using atomic force microscopy (AFM) as a function of time. The PMMA is 40 nm thick and ambient temperature is 300 K. The substrate is 90 nm $SiO_2$ on Si. After 1, 5 and 10 s the resulting nanotrench widths are 61, 80, 81 nm, respectively (at constant heating power of 0.31 mW/μm in the CNT). These time scales are much longer than the thermal time constant of the CNT + PMMA system (up to hundreds of nanoseconds [1, 2]), suggesting that viscous flow plays a role in the trench formation process.

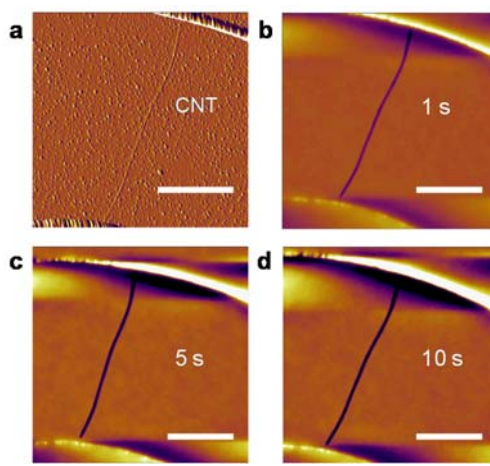

**Figure S4** (a) AFM image of CNT on SiO$_2$ before coating with PMMA. (b-d) AFM images of nanotrench in PMMA after the CNT device is heated with 0.31 mW/µm for different time durations (1, 5 and 10 seconds) as listed. The metallic CNT has a length of 2.6 µm and a diameter of ~2 nm. Scale bars are 1 µm.